\documentclass[a4paper,10pt]{article}
\usepackage[english]{babel}
\usepackage[utf8x]{inputenc}
\usepackage[numbers]{natbib}
\usepackage{amsthm,amsfonts,amsmath,amssymb}
\usepackage{mathptmx,mathrsfs,mathtools}
\usepackage{bm}
\usepackage{cancel}
\usepackage{hyperref}
\usepackage{color,xcolor}

\title{Intuition and importance of feedback control through laboratory experiences}
\author{Aldo Jonathan Munoz-Vazquez}
\date{}

\begin{document}
\maketitle

{\it\small
\begin{minipage}[c]{0.95\textwidth}
Abstract.-- 
This work aims to raise awareness among engineering students from different disciplines on the importance of feedback control. The proposal consists in comparing the performance of different control strategies in a laboratory session, considering Matlab/Simulink simulations of the non-linear pendulum model. First, students attempt to make the pendulum stop at unstable equilibrium by controlling the torque input with a joystick connected to the computer via an Arduino board. Different friction scenarios are considered for students to explore the dissipation in the system response. Then, as a second task, the Arduino is used to introduce the position reference, while students implement different control strategies, such as Bang-Bang, PID (proportional-integral-derivative) and FPID (fractional PID) controllers, analyzing the system response by inspecting the signals in a scope and in a 3D animated model. The dynamic model results as an application of the laws of rotational motion, and the control methods are explained from an intuitive point of view, focusing on the meaning and motivation of the control actions, with the intention to develop intuition about PID and FPID control methods. \\[2mm]
\end{minipage}
}

\section{Introduction}
Control systems are preponderant in the design and implementation of modern advanced technologies and industrial processes that require innovative solutions \cite{antsaklis1999report}, creating a growing demand for highly skilled engineers who can design and develop reliable, robust and autonomous systems.
For this reason, control courses should incorporate a practical curriculum, with laboratory sessions that focus on numerical simulations and physical experiments, including the case of real-time simulations with hardware-in-the-loop and computer-based animated models to allow realistic visualizations, \cite{stefanovic2011labview}.
These kinds of laboratory practices have a positive influence on learning skills, as students retain more knowledge by actively participating in hands-on activities \cite{muske2003introductory,dittmar2016lab}, a fact that is particularly important in control engineering \cite{edward2002role}.

This paper recognizes the importance of acquiring laboratory experience, as it develops intuition and understanding of control principles, motivating the study of analytical concepts from a formal perspective.
Engineers must be knowledgeable beyond theory (physical principles and mathematical equations), which requires performing laboratory practices from an educational perspective \cite{feisel2005role}. 
The key point of laboratory experience in control engineering consists in understanding the importance of control systems and gaining insight about their implementation. In particular, this work is dedicated to the case of feedback control, where control systems are based on measuring system signals to produce adequate actions of control, and thus affecting the evolution of the system dynamics.

Control theory is an integral part of various engineering curricula; for that reason, a successful activity on control laboratory should illustrate different aspects of engineering, while being comprehensive and engaging, showing the importance of theoretical ideas, relevance to different real-world engineering problems, and providing adequate visual information. These activities should also prioritize safety, be affordable and easy to understand, \cite{kheir1996control}.

Successful engineering education depends on a balance between theory and practice \cite{krivickas2007laboratory}. For instance, classroom simulations to reinforce studied topics and laboratory sessions to allow additional practice and acquisition of experimental skills.
The idea of incorporating simulations as part of a laboratory activity also provides additional flexibility in remote work on the concepts studied, \cite{fabregas2011developing,ionescu2013remote,monzo2021remote}. 
Simulations emulate ideal situations that are difficult or impossible to achieve in practice due to physical and technological limitations. Furthermore, simulations are used to test newer control designs, as they provide control performance closer to theory, unlike the case of physical experimentation, where several variables and phenomena are not considered while modeling the system dynamics.

It is acknowledge that simulations are not a replacement for experiments, but as a first step, they allow students to gain intuition about physics and engineering, as well as about the implementation of different methodologies.
Lack of immersion in remote and virtual laboratories has been noted, \cite{da2023extended}, and one problem is the visualization of the system through a display, which limits the ways in which students interact with the process.
In this work, it is proposed to use the Arduino platform to emulate the control of a real system, where different command signals are introduced to the simulation, by attaching hardware to the board. 
In \cite{bashir2019effectiveness}, it was shown that students found the Arduino platform a helpful tool to improve their knowledge of circuits and programming, while they also felt committed to the course activities.
The results of \cite{liu2020implementation} showed that the integration of Matlab/Simulink for control topics helped students understanding  the principles and concepts studied during the course, encouraging them to study more and independently.
Matlab/Simulink and Arduino can be interfaced to produce realistic simulations that have the ability to interact with hardware, allowing the user to enter data while the simulation is running.
Successful control implementations of Simulink and Arduino can be found in \cite{uyanik2018low,yumurtaci2020liquid,ma2021simulation}.
An additional advantage of a simulation in Simulink and Arduino is that the plant dynamics can be potentially replaced by a real world physical system, collecting data from sensors, and sending control signals to the actuators, which are produced by the Simulink blocks.

The contribution of this paper is stated as follows:\\[-2mm]

{\it 
The development of a laboratory practice for engineering students, so that they understand the importance of closed-loop control, and at the same time, learn different control methods, establishing strong intuitive connections with theoretical concepts.\\[-2mm]
}

It is recommended that student observations and conclusions be recorded in surveys to analyze the effectiveness of the proposed practice and implement convenient changes at the discretion of the instructor. 

The remaining of this paper is organized as follows: Next section describes the proposed activities in the laboratory practice. Section \ref{Sec:Goals} exposes the educational goals related to this practice. Section \ref{Sec:Model} presents the modeling of the pendulum system and briefly described the open-loop strategy. Section \ref{Sec:Control} presents the studied control strategies. Finally, Section \ref{Sec:Conc} provides discussions and future work. 

\section{Methodology}

The proposed laboratory activity is organized as follows:

\begin{itemize}
    \item The dynamic model of the pendulum is obtained during lecture, extending Newton's laws of motion to the rotational case. During the same session, students discuss practical implications of these models, and elucidate different strategies for control purposes.
    \item Students incorporate the equation of motion of the pendulum system into a Matlab function block in Simulink. They code a linearized version around the stable equilibrium, and perform comparisons for different values of the initial condition. The visualization is carried out by scoping signals and by a 3D animated model.
    \item Students try to stop the pendulum at the unstable equilibrium using a joystick that is connected to an Arduino board, entering torque signals to the pendulum. Different scenarios are considered, with different friction values. Students draw their conclusions on the effect of changing the system viscosity.
    \item Students enter a reference for the angular position. At this point, they are aware of the need for a control strategy to produce a torque signal that depends on both the desired reference and the pendulum angle. Two main options are discussed, namely: {\it i)} bang-bang control and {\it ii)} P (proportional) controller, considering the cases of zero, low, moderate and intense friction. The purpose is that students realize the effect of varying dissipation in control performance.
    \item Understanding that friction is a system parameter, which cannot be modified in most of the cases, students will artificially inject friction to the system, this is, through the action of a D (derivative) control. D-action alone is useless, as it does not account for the angular position error. In this sense, students implement the PD (proportional-derivative) controller. The intuitive approach is that the P-action tries to improve accuracy, relaying on the current error value, and the D-action tries to improve stability, while predicting the next instantaneous error value.
    \item Students are introduced to the concept of integral (I) control, which induces interesting effects, such as robustness against constant disturbances, providing additional parameter to improve the closed-loop performance. Then, students test PI, PD and PID controllers.
    \item The D-action is sensitive to measurement noise and the I-action affects the relative stability of the closed-loop system during the transient period. Then, students learn about the existence of fractional derivatives and integrals, implementing the FPID, and comparing its performance with respect to the other discussed control methods. For comparisons, the reference is introduced using a sine wave reference; input and measurement noises are also introduced, and a non-constant input disturbance is considered.
\end{itemize}

\section{Educational Goals}\label{Sec:Goals}
The proposed laboratory experience aims to replicate the complexity of real-world experiments by integrating the simulated model of a pendulum system with user-inputs through an Arduino board. This is beneficial for junior and senior students, as they learn from both theoretical derivations and hands-on activities, interacting with the system, and analyzing the effects of changing system parameters and control strategy.
The use of Arduino with Simulink imposes some technical challenges that must be resolved by students, introducing some technical concepts such as serial communication, communication channels, baud-rate, etc. 
By implementing a numerical model in Simulink, students also explore different numerical algorithms (Euler, Trapezoidal, Runge-Kutta, etc.), learn about sampling time, and are aware of the existence of fixed-step and variable-step integration methods. They also learn about integrating 3D animated models in Simulink, defining rotations and translations. 

Overall, this laboratory practice considers the following educational goals:

\begin{itemize}
\item 
{\it The first educational goal:} Learn physical principles for modeling engineering systems and describing physical concepts in mathematical terms, that is, in the form of ordinary differential equations.
\item 
{\it The second educational goal:} Learn to solve dynamic models using numerical methods in Simulink. This consists in identifying and algebraically solving the highest-order derivative term in the system dynamics, and finding and feeding back the other variables by means of numerical integration with built-in initial conditions.
\item 
{\it The third educational goal:} Learn several simulation tools and hardware devices. The use of a joystick consisting of a potentiometer, which sends an analog value between 0V and 5V to the Arduino, is considered; then, this value is converted into a number between 0 and 1023 by means of the 10-bit ADC. Students then use appropriate blocks to transform that number into the correct format for a torque in the range $[\tau_{\min},\tau_{\max}]$, or an angular position reference in $[r_{\min},r_{\max}]$.
\item 
{\it The fourth educational goal:} Understand the importance of feedback control, acquiring knowledge on the application of conventional formulations, such as P, PD, PI and PID schemes, exploring more advanced alternatives, such as the FPID controller. It is desirable that students be motivated to study all these topics in a formal way, enrolling in control engineering and control theory courses.
\end{itemize}

\section{Modeling and Open-loop Control}\label{Sec:Model}
\begin{figure}[t!]
    \centering
    \includegraphics[height=3.5cm]{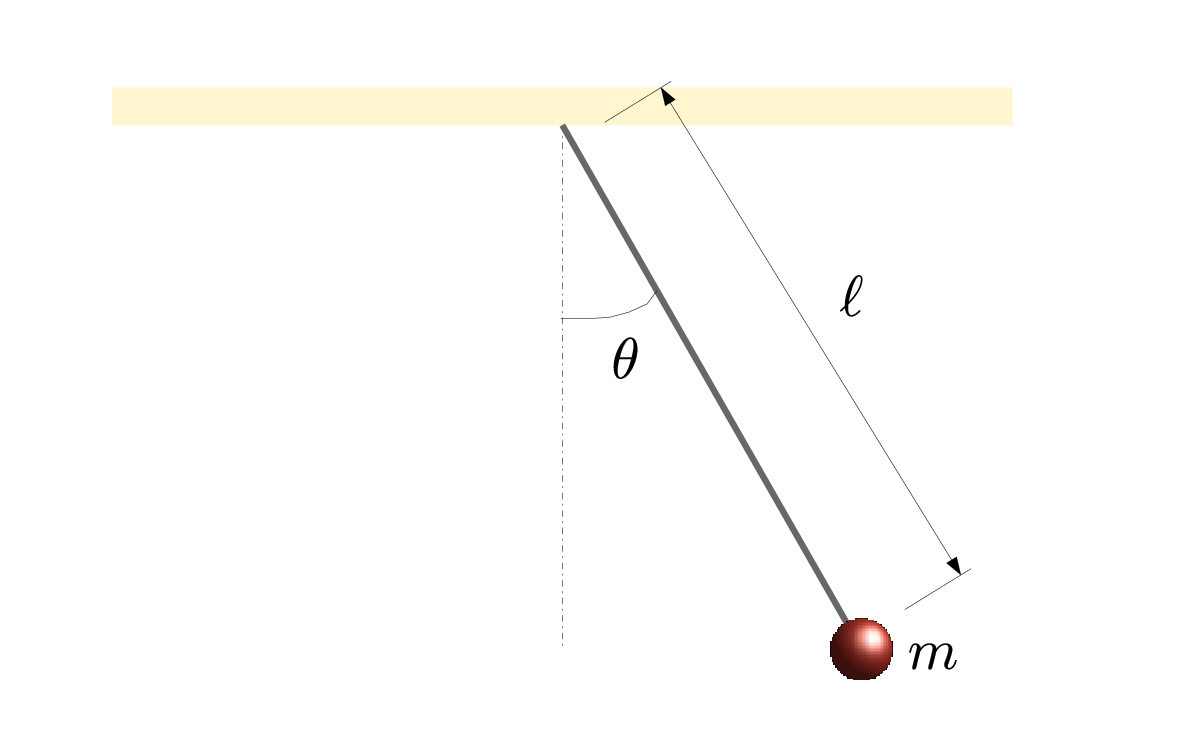}
    \includegraphics[height=3.0cm]{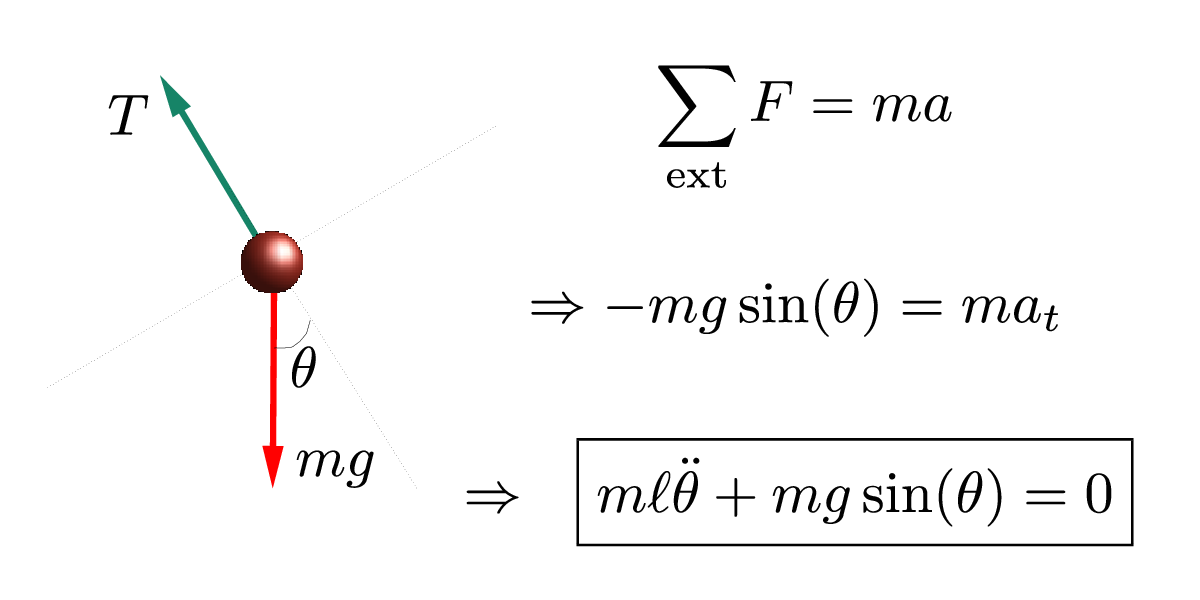}
    \caption{Simple pendulum system.}
    \label{fig:pendulum}
\end{figure}

The system under consideration is the simple pendulum system of Fig. \ref{fig:pendulum}. For a first modeling attempt, the following assumptions are made: 
\begin{itemize}
    \item The rod is mass-less and infinitely rigid.
    \item There is no friction or any other dissipation effects.
    \item The mass at the end of the rod is condensed in a single point, with no volume.
    \item The motion is perfectly constrained into a plane.
\end{itemize}
Under these conditions, it is evident that a single variable, the angle $\theta$, is enough to describe the system configuration. Then, the pendulum system is of 1-DoF (degrees of freedom).

The assumption that the rod is infinitely rigid causes the tension force to be completely canceled out by the sum of the centrifugal force and the projection of the gravitational force projected onto the line joining the mass and the pivot. This is particularly important for estimating required tensile strength in a real system, and selecting the adequate material to withstand various effects, keeping the system operating under adequate conditions.

The particle motion occurs in a circumference, and its velocity vector is perpendicular to the rod. The equation of motion for the ideal system is  
\begin{equation}\label{Eq:nonlinear}
    m \ell \ddot\theta(t) + mg\sin(\theta(t)) = 0,
\end{equation}
for $t$ the time parameter in [s], $m$ mass of the pendulum in [kg], $\ell$ the length of the rod in [m], $g$ the gravity constant in [m$/$s$^2$], and $\theta$ the angle with respect to the vertical line in [rad]. The above equation can be reduced to $\ddot\theta(t) + \frac{g}{\ell}\sin(\theta(t)) = 0$, and can be used to estimate the value of $g$, using the approximation
\begin{equation}
    g \approx \frac{4\pi^2 \ell}{T^2}, 
\end{equation}
for $T$ the period of oscillation in [s]. something that students could try during the proposed activity.
The linear pendulum model 
\begin{equation}\label{Eq:linear}
    \ddot\theta(t) + \frac{g}{\ell}\theta(t) = 0
\end{equation}
is very popular in control literature to design linear structures, demonstrating stability properties of the closed-loop system in the frequency domain. The linear model relies on $\sin(\theta)\approx\theta$ for small values of $|\theta|$. The students compare models \eqref{Eq:nonlinear} and \eqref{Eq:linear} for different initial conditions, reporting their observations and conclusions. Then, the instructor discusses on the validity of these models in real-world scenarios.

Model \eqref{Eq:nonlinear} is appropriate for visualizing the solution of the pendulum equation. However, to understand the effect of external inputs, such as actuating the pendulum by including a DC-motor in the pivot, one needs to consider the Euler law for rotational motion, which states that the sum of the torques about a fixed point equals the system moment of inertia about that point times the angular acceleration of the rod. One should be aware that the tension force provokes no torque about the pivot, while the only force causing a rotational motion is the projection of the gravity force onto the line that is perpendicular to the rod, producing the torque $-mg\ell\sin(\theta)$. Thus, the sum of torques about the pivot renders the equation of motion 
\begin{equation}\label{Eq:rotational}
    m \ell^2 \ddot\theta(t) + mg\ell\sin(\theta(t)) = \tau,
\end{equation}
for $\tau$ the torque delivered by the DC-motor actuator. It is also possible to define $J = m \ell^2$, as the moment of inertia of the particle mass with respect to the pivot point. 
Moreover, with the intention of making model \eqref{Eq:rotational} more realistic, one includes the effect of dissipation forces, resulting
\begin{equation}\label{Eq:rotational2}
    J \ddot\theta(t) + b \dot\theta(t) + mg\ell\sin(\theta(t)) = \tau,
\end{equation}
for $b$ the viscous friction coefficient in [N$\cdot$m$\cdot$s/rad].

\begin{figure}[t!]
    \centering
    \includegraphics[width=10cm]{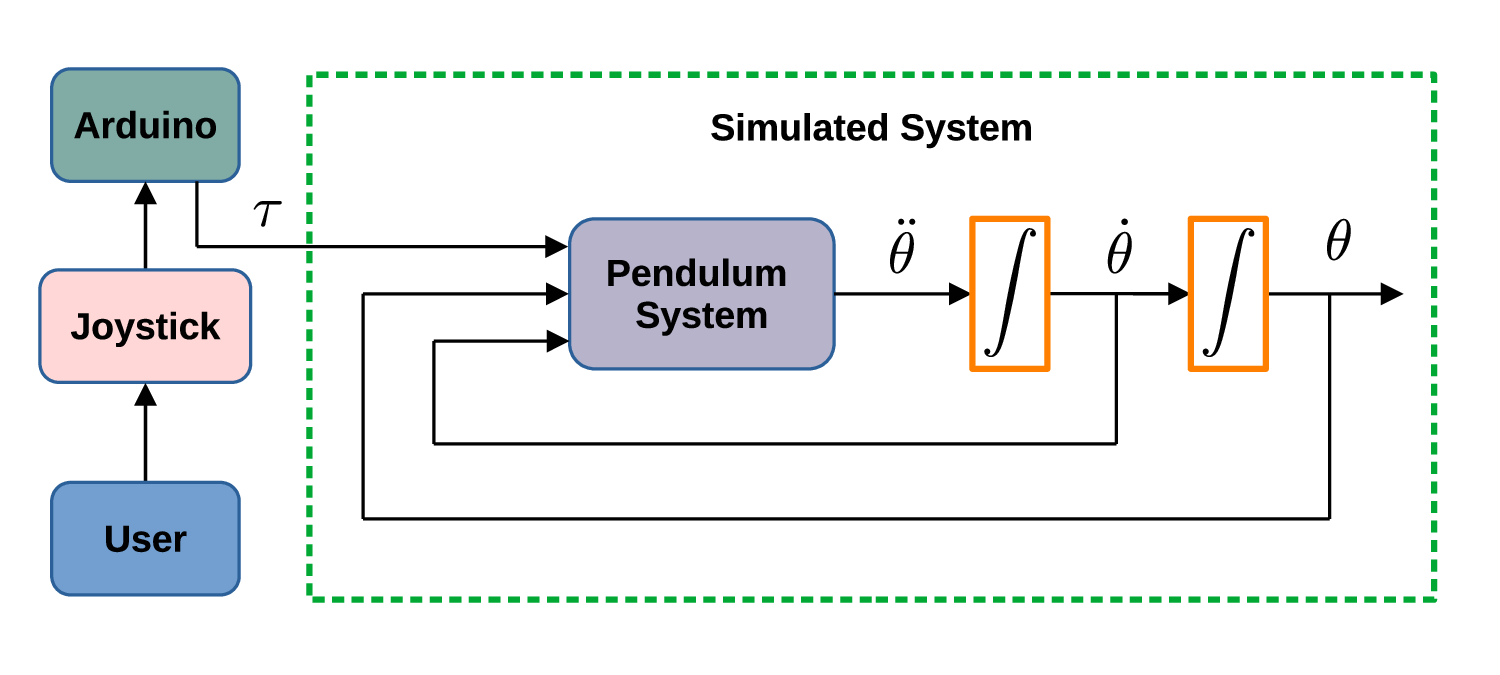}
    \caption{Open-loop control.}
    \label{fig:open}
\end{figure}

\begin{figure}[b!]
    \centering
    \includegraphics[width=9cm]{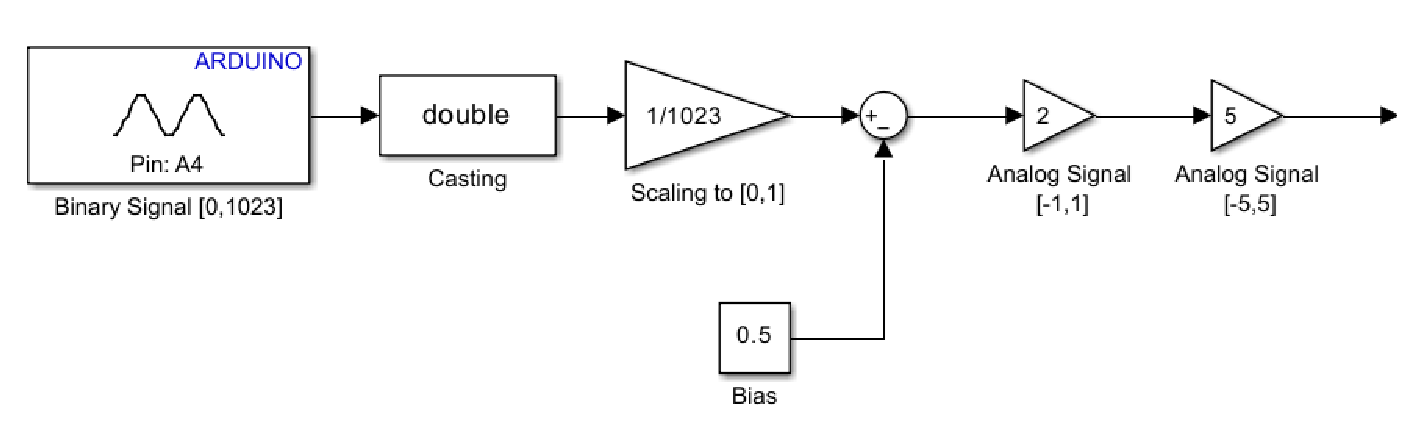}
    \caption{Torque signal in the required range.}
    \label{fig:torque}
\end{figure}

The students are asked to enter $\tau$, in the range $[-5,5]$N$\cdot$m, to stop the pendulum at the unstable point $\theta = \pi$. {\it The pendulum must remain static after reaching the desired configuration, and without the need of additional human intervention.}
The open-loop system is depicted in Fig. \ref{fig:open}.
The torque signal is adjusted into an analog signal using the process shown in Fig. \eqref{fig:torque}. 
System parameters in \eqref{Eq:rotational2} are $\ell = 0.20$m, $g = 9.81$ m$/$s$^2$, $m = 0.2$kg, and are considered as constants. Different friction cases are considered $b \in \{0.0,~0.1,~0.5,~1.0\}$N$\cdot$m$\cdot$s$/$rad.
The simulation runs on the fourth-order Runge-Kutta method, where the sampling time is adjusted trying to emulate the real-time clock signal.

Students are invited to answer (anonymously) the following questions on a scale from 1 to 5, where a value = 1 represents low and a value = 5 represents high.
\begin{enumerate}
    \item Were the instructions clear? That is, did you understand what you were asked to do during this section of the practice?
    \item After completing this activity, do you feel more confident in understanding the pendulum model?
    \item Do you think this activity contributed to your knowledge about dynamical systems and mathematical modeling?
    \item Do you believe that this activity provides important tools for your future career as an engineer?
    \item Does increasing friction help stabilize the pendulum system at the reference point?
\end{enumerate}

\section{Closed-loop Control}\label{Sec:Control}
\begin{figure}[t!]
    \centering
    \includegraphics[width=11cm]{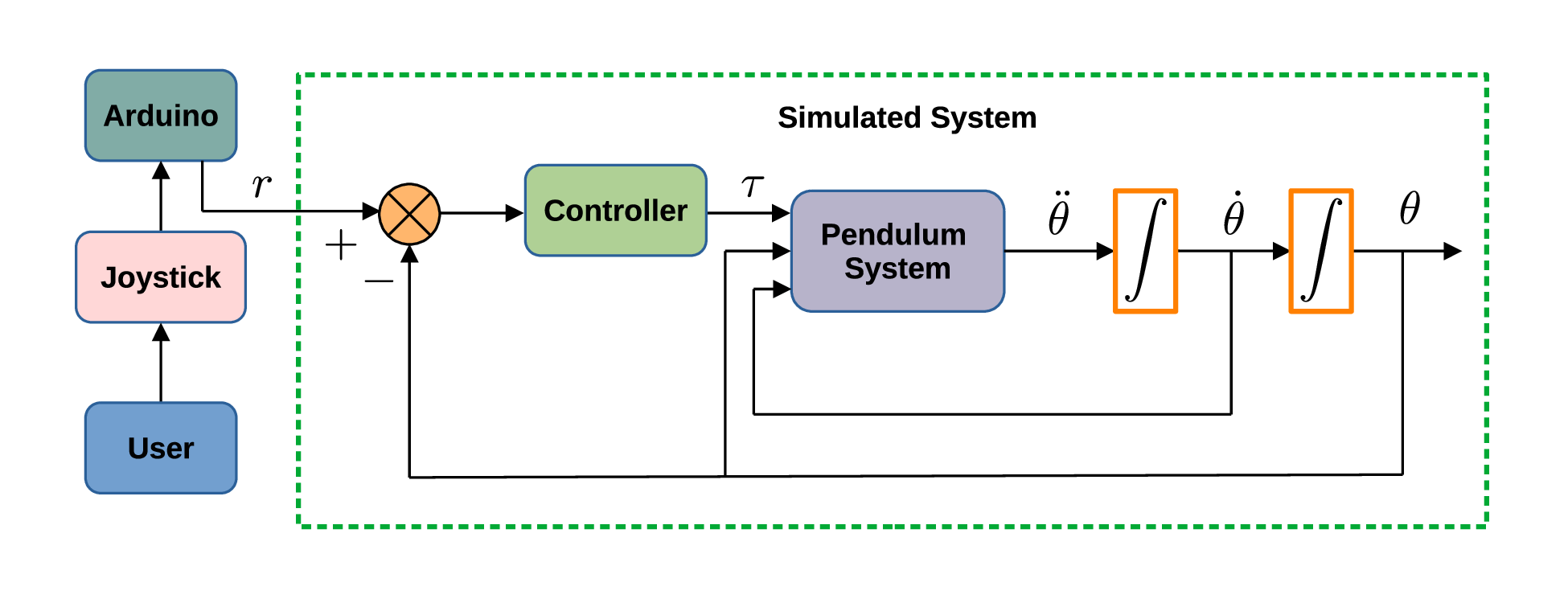}
    \caption{Open-loop control.}
    \label{fig:closed}
\end{figure}

In this section of the practice, students are asked to modify the simulation to include the joystick signal as an angular position reference. The goal is to design a control strategy such that the pendulum angle follows the reference dictated by the user. The diagram of the modified simulation is shown in Fig. \ref{fig:closed}. The reference, coming from the user, must be conditioned to obtain the signal within the required limits as $r \in [-\pi,\pi]$. 
The obtained reference signal is low-pass filtered, with the purpose of removing noisy components and/or discontinuities.
The correct signs at the comparison point are inferred by considering that the torque control should drive the output $\theta$ towards the reference.

\subsection{Bang-Bang Control}
The first control method to be consider is the bang-bang scheme, which provides a commuting control action depending on the signum of the error, this is,
\begin{equation}\label{Eq:bang-bang}
    \tau_{1} = \left\{
    \begin{matrix}
        \tau_{\rm max} & \mbox{for} & \theta < r \\
        0 & \mbox{for} & \theta = r \\
        -\tau_{\rm max} & \mbox{for} & \theta > r
    \end{matrix}
    \right.
\end{equation}
with $\tau_{\rm max}$ the maximum (allowed) torque that can be exerted by the actuator. This control strategy can be also written as 
\begin{equation}
    \tau_{1} = \tau_{\rm max} \mbox{sgn}(r - \theta).
\end{equation}
An energy-based analysis for the system with $\tau =\tau_1$ leads to 
\begin{equation}\label{Eq:energy}
\dot E(\theta,\dot\theta) =  - b \dot\theta^2(t) - \tau_{\rm max} \dot\theta \mbox{sgn}(\theta - r),
\end{equation}
for $E(\theta,\dot\theta) = K(\theta,\dot\theta) + U(\theta)$ the total energy, with $K(\theta,\dot\theta) = \frac{1}{2}J\dot\theta^2$ the kinetic energy and $U(\theta) = - mg\ell\cos(\theta)$ the potential energy. 
For constant reference $r$, one can recognize that $\frac{d}{dt} |\theta - r| = \dot\theta \mbox{sgn}(\theta - r)$, leading to 
\begin{equation}\label{Eq:energy_a}
\dot E_a(\theta,\dot\theta) =  - b \dot\theta^2(t),
\end{equation}
for $E_a(\theta,\dot\theta)  = E(\theta,\dot\theta) + \tau_{\rm max}|\theta - r|$ the augmented energy function. The above means that the augmented energy is monotonically decreasing as long as $b>0$ and $\dot\theta \neq 0$. This implies that the system trajectories $(\theta(t),\dot\theta(t))$ converges to an invariant set where $\dot E_a(\theta^*,\dot\theta^*) \equiv 0$, this is, where $\dot\theta^* \equiv 0$ and $\ddot\theta^* \equiv 0$, for $\equiv$ meaning ``equivalently equal to'' or ``in average''.
These conditions lead to infer that the system evolves towards a static equilibrium, which can be found from
\begin{equation}\label{Eq:equil}
    J \cancel{\ddot\theta^*} + b \cancel{\dot\theta^*} + mg\ell\sin(\theta^*) \equiv \tau_{\rm max} \mbox{sgn}(r-\theta^*).
\end{equation}
From the consign $mg\ell\sin(\theta^*) \equiv \tau_{\rm max} \mbox{sgn}(r-\theta^*)$, it can be seen that for $\tau_{\rm max} > mg\ell$, the only possible solution is $\theta^* = r = \pi$. For the case of $r\notin\{0, \pi\}$ a similar analysis is possible, but requiring to consider the set-valued evaluation $\tau_1(r) \in [\tau_{\rm max},\tau_{\max}]$, leading to $\tau_1(r)\equiv mg\ell\sin(r)$. In this latter case, the solutions of the system dynamics are understood in the sense of Filippov.

\subsection{Proportional Control}
From the above experience, students learn that bang-bang control is unsuitable in several applications, and decide to move to a smoother alternative, such as the P control, which injects a control torque that is proportional to the error, this is,
\begin{equation}\label{Eq:P}
    \tau_{2} = k_p (r - \theta),
\end{equation}
for proportional gain $k_p>0$. It is interesting to notice that the same (at least mathematically) dynamics could be obtained by including a rotational spring, with rigidity $k_p$, at the point $\theta = r$; thus, the control torque acts as a restitution force. 

The energy-based analysis used above, but for $\tau =\tau_2$, leads to 
\begin{equation}\label{Eq:energy2}
\dot E(\theta,\dot\theta) =  - b \dot\theta^2(t) - k_p \dot\theta (\theta - r),
\end{equation}
where, for constant reference $r$, one has $\frac{d}{dt} \left\{ \frac{1}{2}(\theta-r)^2 \right\} =  \dot\theta (\theta-r)$, which produces
\begin{equation}\label{Eq:energy_a2}
\dot E_a(\theta,\dot\theta) =  - b \dot\theta^2(t),
\end{equation}
with $E_a(\theta,\dot\theta)  = E(\theta,\dot\theta) + \frac{1}{2}k_p(\theta-r)^2$ the augmented energy function. As in the previous case, for friction coefficient $b>0$, the trajectories of the system converge to an invariant set, with 
\begin{equation}\label{Eq:equil2}
    J \cancel{\ddot\theta^*} + b \cancel{\dot\theta^*} + mg\ell\sin(\theta^*) = k_p(r-\theta^*).
\end{equation}
It can be noticed that,  for $r = \pi$ and $e^* = \pi-\theta^*$, one has that $\sin(e^*) =  \sin(\theta^*)$. Then, the equilibrium equation is $mg\ell\sin(e^*) = k_p \, e^*$, and produces the nontrivial solution $e^* =0$ for arbitrary values of $m$, $\ell$ and $k_p$.
In contrast to the case of the Bang-Bang controller, for a reference $r\notin\{0,\pi\}$, the equation $mg\ell\sin(\theta^*) = k_p(r-\theta^*)$ sustains for values $\theta^*\neq r$, which can be tested during the simulation activity.
It can be seen that the proportional gain helps to increase the accuracy of $\theta^* \approx r$ as well as speeding up the convergence rate, 
at the expense of a larger control signal.

There is a shared problem with controllers \eqref{Eq:bang-bang} and \eqref{Eq:P}, and is that they are only able to enforce stable regulation of the unstable equilibrium, $r = \pi$, as long as $b>0$. For the case of $r$ is not a multiple of $\pi$, only the bang-bang controller \eqref{Eq:bang-bang} is able to induce convergence $\theta\to r$ as $t\to\infty$, but this controller is discontinuous, and unfeasible for implementation in most of mechanical systems, as it induces self-sustained oscillations in the system response. 

The first problem can be solved by introducing a derivative action, which acts as a dissipation, while the second problem, for the case of regulation, that is, when $r$ is a constant reference, is solved by including a compensation of the gravitational effect, either at the desired point or for the entire range of motion.  A powerful alternative consists in including an integral action, providing the system with robustness capabilities against constant disturbances. The tracking case, when reference $r =r(t)$ is a time variant function, is out of the scope of this paper; nonetheless, the studied controllers are capable of perfoming acceptable tracking performance.

\subsection{PD Control}

In this case, the control torque is computed as
\begin{equation}\label{Eq:PD}
    \tau_3 = k_p(r - \theta) + k_d (\dot r - \dot\theta).
\end{equation}
with feedback gains $k_p$, $k_d>0$, which for a constant reference $r$, becomes into
\begin{equation}\label{Eq:PD_double}
    \tau_3 = k_p(r - \theta) - k_d \dot\theta.
\end{equation}
The form in \eqref{Eq:PD_double} is preferred over \eqref{Eq:PD}, as the reference $r(t)$ could contain any sort of discontinuities. Nevertheless, in this paper $r(t)$ is obtained by means of an Arduino board, and a low-pass filter is considered to remove any noise and/or discontinuities from the signal.

The same analysis as in the case of the P controller leads to
\begin{equation}\label{Eq:energy3}
\dot E_a(\theta,\dot\theta) =  - (b + k_d) \dot\theta^2(t),
\end{equation}
with augmented energy function $E_a(\theta,\dot\theta)  = E(\theta,\dot\theta) + \frac{1}{2}k_p(\theta-r)^2$. In this case, even for $b = 0$, one has that, for $r$ being a multiple of $\pi$, one gets $\theta\to r$ as $t\to\infty$.

\subsection{PID Control}
It can be seen that the PD controller enforces $\theta\to r$ for $k_d>0$, whenever $r$ is a multiple of $\pi$. However, in other case, when $r\notin\{0,\pi\}$, it is necessary to consider additional tools. For instance, including an integral action of control allows to enforce $\theta\to r$, for arbitrary constant $r$, and even in the presence of constant input disturbances. 
The studied PID controller is 
\begin{equation}\label{Eq:PID}
    \tau_4 = k_p(r - \theta) + k_i \int_0^t [r(t') - \theta(t')] dt' + k_d (\dot r - \dot\theta).
\end{equation}
or in the case of a constant reference $r$,
\begin{equation}\label{Eq:PID_double}
    \tau_4 = k_p(r - \theta) + k_i \int_0^t [r - \theta(t')] dt' - k_d \dot\theta.
\end{equation}
The equation of motion for the pendulum system in closed-loop with the PID controller \eqref{Eq:PID_double} is
\begin{align}
\begin{split}
        J \ddot\theta + b \dot\theta + mg\ell\sin(\theta) &=  k_p (r-\theta) - k_d\dot\theta + \sigma,\\
        \dot\sigma &= - k_i (r-\theta),
\end{split}
\end{align}
whose equilibrium, where all derivatives are set at zero, can be determined from  
\begin{align}
\begin{split}
        mg\ell\sin(\theta^*) &=  k_p (r-\theta^*) + \sigma^*,\\
        0 &= - k_i (r-\theta^*),
\end{split}
\end{align}
producing, simultaneously, $\theta^* = r$ and $\sigma^* = mg\ell\sin(r)$. In other words, the integral action compensates gravitational effects evaluated at the equilibrium. The energy-based analysis is more involved in this case due to the presence of crossed terms in the derivative of the energy function. Nevertheless, for the interested reader, the derivation can be computed using the method proposed in \cite{kelly1995tuning}. It is important to mention that large values for the integral gain $k_i$ usually lead to catastrophic consequences, such as instability. Something that can be tested in a simulation activity.

\subsection{FPID Control}
It was made evident that the integral action increases robustness and accuracy, but at the price of increasing oscillations during the transient period. The addition of measurement noise also makes it clear that the derivative action is sensitive to high-frequency effects. 
As a way to alleviate those drawbacks of PID control, the following alternative, known as FPID control \cite{podlubny1999fractional}, is considered,
\begin{equation}\label{Eq:FPID}
    \tau_5 = k_p(r - \theta) + k_i I^{\lambda} (r - \theta) + k_d D^{\mu}( r - \theta),
\end{equation}
for $I^{\lambda}$ and $D^{\mu}$ the fractional integral and derivative operators of order $\lambda$ and $\mu\in(0,1)$, respectively, and constant gains $k_p$, $k_d$ and $k_i>0$. 
The formal analysis of the FPID is beyond  the scope of this paper, but it is beneficial for students to consider the existence of additional and more advanced control tools. 
This also constitutes a motivation to explore alternative ideas, shaping the technologies of the future.

\subsection{Students Evaluations}
Students are invited to answer the following questions on a scale from 1 to 5, where a value = 1 represents low and a value = 5 represents high.
\begin{enumerate}
    \item To what extent did you understand what you were asked to do during this section of the practice?
    \item After completing this activity, do you feel more confident in understanding the purpose of feedback control?
    \item Do you think this activity helps to increase your knowledge of control systems?
    \item Do you feel motivated to learn about control systems in the future?
    \item Do you believe that this activity provides important tools for your future career as an engineer?
\end{enumerate}

The next questions aim to evaluate different aspects of the studied controllers from a qualitative point of view.
\begin{itemize}
    \item Which of the controllers studied did you find most difficult to understand?
    \item Regardless of the difficulty, which control strategy do you consider the best, according to performance and parameter tuning?
    \item In the presence of noise, what controller do you consider the best, in terms of tracking accuracy?
    \item In the presence of low friction, what controller do you consider best, in terms of tracking accuracy?
    \item In the presence of moderate friction, what controller do you consider best, in terms of tracking accuracy?
    \item Suppose you are designing a new technological application that requires some control implementation, which algorithm would you like to use? (you can select more than one option),
    \item What do you believe is more important when designing a control scheme: robustness against disturbances, sensitivity to measurement noise, smoothness of the control signal, regulation/tracking accuracy?
\end{itemize}

Students are also invited to write down their opinions, perspectives, reflections and suggestions.

\section{Discussions and Future Work}\label{Sec:Conc}
Although the proposed work presents some limitations, as the dynamic model is considered as an approximation since some physical constraints are not included in the model, such as actuator limitation, sampling, input/measurement delay, quantification, etc., this practice has the sole purpose of serving as a first and friendly approach to control systems for junior and senior engineering students, whose main objective is to foster motivation and engagement in upcoming lectures and more specialized courses in control theory and applications.
During this laboratory activity, students will gain greater insight into the need for advanced mathematical tools to model and control physical processes in engineering practice.
Students are also exposed to different technological concepts, which make them aware of the usefulness of electronics, programming and communications in designing high-end engineering solutions.

Future work is considered to explore laboratory experiences where students design different systems that provide solutions from engineering and technology perspectives to well-defined real-world problems, relying on concepts related to robotics, system dynamics, controls and applied mathematics. Students will be expected to model and design mechanical structures and 3D print different parts required to implement their designs.

\section*{Conflict of interest}
The author of this paper does not have any conflict of interest regarding the publication of this paper.

\bibliographystyle{unsrt}

\bibliography{example}

\end{document}